# Achieving sub-diffraction imaging through bound surface states in negative-refracting photonic crystals at the near-infrared

R. Chatterjee[1], N. C. Panoiu[2], K. Liu[1], Z. Dios[1], M. B.Yu[3], M. T. Doan[3], L. J. Kaufman[1], R. M. Osgood[1], and C. W. Wong[1]

[1]*Columbia University, New York, NY 10027, USA*

[2]*University College London, Torrington Place, London WC1E 7JE, UK*
[3]*The Institute of Microelectronics, Singapore 117685, Singapore*

We report the observation of imaging beyond the diffraction limit due to bound surface states in negative refraction photonic crystals. We achieve an effective negative index figure-of-merit [-Re($n$)/Im($n$)] of at least 380, ~125× improvement over recent efforts in the near-infrared, with a 0.4 THz bandwidth. Supported by numerical and theoretical analyses, the observed near-field resolution is 0.47λ, clearly smaller than the diffraction limit of 0.61λ. Importantly, we show this sub-diffraction imaging is due to resonant excitation of surface slab modes, allowing refocusing of non-propagating evanescent waves.

PACS: 78.20.Ci, 41.20.Jb, 42.70.Qs, 78.67.-n

Recently it has been suggested theoretically that left-handed metamaterial (LHM) lens can recover evanescent waves, and information in both propagating and evanescent fields can be reconstructed into a perfect image [1]. Metal-based left-handed metamaterials containing resonant metallic microstructures operating at optical frequencies have recently been demonstrated remarkably to possess negative refraction [2,3] negative magnetic permeability [4], and even introduced as superlenses [5]. Other approaches, such as materials with indefinite permittivity and permeability tensors [6], domain twin structures [7] and dielectric metamaterials in the mid-infrared [8] and near-infrared [9], also suggest interesting potential towards negative refraction. An alternative experimental approach is to use dielectric-based photonic crystals (PhCs) where optical losses are considerably smaller and whose dispersion properties can be engineered so that at specific frequencies they possess negative refraction.

While theoretical studies have predicted subwavelength imaging in PhC-based LHMs at optical and near-infrared frequencies [10,11], an experimental demonstration of this phenomenon at these frequencies is still critically lacking, primarily due to the significant challenges in device nanofabrication and measurements. Here, by using a two-dimensional (2D) PhC slab, which emulates an effective negative refractive index over a specific frequency range, we achieve the first near-field observation of sub-diffraction limit imaging at near-infrared frequencies in a photonic structure with a large negative index figure-of-merit (FOM) of at least 380. This FOM is about 125× improvement over recent experimental efforts [2] and more than 15× better than the best theoretical efforts [12]. Importantly, we also clearly observe, for the first time, that this sub-diffraction limit imaging is due to resonant excitation of surface slab modes, permitting refocusing of the non-propagating evanescent near-fields.

The possibility of left-handed behavior of wave propagation in PhC is intimately related to the existence of photonic modes with negative phase index, *i.e.* modes for which Bloch wavevector $\boldsymbol{k}$ in the first Brillouin zone and the group velocity $\boldsymbol{v}_g$, defined as the gradient of the mode frequency $\omega(\boldsymbol{k})$, $\boldsymbol{v}_g = \nabla_{\boldsymbol{k}}\omega(\boldsymbol{k})$, are antiparallel, *i.e.* $\boldsymbol{k}\cdot\boldsymbol{v}_g < 0$ [11,13]. In particular, negative refraction and sub-diffraction limit imaging in such PhCs have been predicted theoretically [10,11] and demonstrated at microwave frequencies [14,15] and recently in the near-infrared [16], although without obtaining sub-diffraction imaging. Although these phenomena would ideally be studied in PhC whose phase index is isotropic, such as near $\Gamma$ point, they can be observed even when the phase index shows certain degree of optical anisotropy [14]. This ability to tolerate PhC anisotropy [14] is particularly important for the 2D PhC slab geometries used in our study, since in this case modes with wavevectors $\boldsymbol{k}$ close to the $\Gamma$ point are leaky modes characterized by large optical losses and thus cannot operate in this near-$\Gamma$ region of the $\boldsymbol{k}$-space. Luo *et al.* [10] show theoretically that a mechanism leading to evanescent wave recovery is excitation of bound transverse guiding states (*slab modes* or *surface slab modes*) of the PhC slab, with the effect of an overall resonance divergence of the slab transmission. Our PhC is designed such that, for our wavelengths, the slab mode symmetry only allows excitation of resonant surface slab modes, for sub-diffraction imaging.

The PhC-based material is a hexagonal lattice (*p6m* symmetry group) with holes etched into a silicon film of thickness $t = 320$ nm, on a $SiO_2$ insulator substrate (we consider the slab median plane to be the *xz*-plane, with *y*-axis normal to this plane) as shown in Fig. 1a. The normal to the input facet is oriented along the $\Gamma$-M symmetry axis of the crystal. The fabricated PhC slab has a lattice constant $a = 433.5$ nm and a radius $r = 119.36$ nm ($r = 0.2753a$). Vertical optical confinement is obtained through total internal reflection, with air-cladding (top) and buried silicon oxide (bottom). Our subwavelength source is a tapered $765 \times 320$ nm$^2$ waveguide placed 250 nm away from the PhC slab. At our measurement wavelengths the waveguide supports only two modes, a $E_{11}^x$ ($E_{11}^y$) mode whose electric (magnetic) field is predominantly oriented along the *x*-axis. A collection-mode near-field scanning optical microscope (NSOM) obtains the optical field in the PhC proximity. The NSOM spatial profile is deconvolved with the simulated waveguide output profile to obtain the non-fiber-perturbed characteristics. Each near-field image measurement is averaged over 27 scans to improve the signal-to-noise.

The PhC slab is designed by employing rigorous three-dimensional (3D) frequency- and time-domain numerical simulations. The modes are strongly confined to the slab, with energy flow parallel to the slab plane. We classify the photonic bands in Fig. 1b according to how the modes transform upon applying the reflection operators $\Sigma_i$ ($i=x,y,z$), and for this we calculated the parameters $\sigma_i$, which quantify the mode parity: $\sigma_i = \int_{BZ} E_k(r) \cdot \Sigma_i E_k(r) dr$, the integration being performed over one unit cell. In our case the PhC slab is not invariant to the reflection operator $\Sigma_y$; however, the computed values of the parity parameter $\sigma_y$ ($\sigma_y \sim \pm 0.98$ to $\pm 0.99$) show that the modes can be meaningfully classified as TE-like and TM-like.

We note that, since our input wavelengths fall in the bandgap of TE-like modes, only TM-like modes are excited. Our designed slab contains two such modes with the same symmetry as the $E_{11}^y$ input waveguide mode; both of these modes are *surface slab modes*. They extend from ~1481 to ~1484 nm (see bottom-right panel in Fig. 1b with *a* = 433.5 nm) and their symmetry properties with respect to $\Sigma_z$ reflections are such that these two modes are a symmetric and antisymmetric superposition of surface modes excited at each air-PhC interface. All other modes are guiding modes of the entire PhC slab and do not have the required symmetry with respect to $\Sigma_z$ reflections and thus cannot be excited. In addition, because of the *y*-axis symmetry of these TM-like modes, they can only be excited by the $E_{11}^y$ waveguide mode for this frequency range. Moreover, of these two TM-like bands only one has the same (*x*-axis) symmetry properties upon $\Sigma_x$ reflection as the $E_{11}^y$ mode does, namely $\sigma_x > 0$; this band is marked in red in Fig. 1b, bottom-left panel. Furthermore, for achieving imaging beyond the diffraction limit, we designed this band to emulate a negative index of refraction, with equi-frequency curves (EFC) that correspond to increasing frequency shrink toward the $\Gamma$ symmetry point (Fig. 1b top-right panel): the effective refractive index varies between $n = -1.64$ at $\lambda = 1480$ nm and $n = -1.95$ at $\lambda = 1580$ nm, determined from $|k| = n\,\omega/c$, for our measurement range. Note, while the EFC shape shows our PhC slab does not have negative refraction for all angles [10], negative refraction is present for a large domain of incident $\boldsymbol{k}$, so that the PhC slab does possess one of the main ingredients necessary to observe sub-diffraction limit imaging.

Figures 2a-d show the near-field measurements. These figures illustrate that while the full-width half-maximum (FWHM) of the input-source waist remains approximately constant at ~1.0 μm (1.5 μm before deconvolution) across all wavelengths, a minimum FHWM waist of the image, of 0.7 μm, *i. e.* 0.47λ, (1.29 μm before deconvolution), is observed at 1489 nm (Fig. 2d). This image size is clearly below the ideal-lens Abbe diffraction limit of 0.91 μm (0.61λ) at $\lambda = 1489$ nm. In comparison, the FHWM waist is 1.9 μm (1.23λ) at 1550 nm. Uncertainty in the FHWM waist is estimated at 0.05 μm. The sub-diffraction imaging bandwidth for this sample is ~3 nm (or 0.4 THz), from the 3-dB width of Lorentzian fit in Fig. 2d. Additional scattering at the top right side of the source in Figs. 2a and 2b is due to fabrication defects, which are not seen in the Fig. 1a. Although these defects increased light scattering at the waveguide input, the large-angle nature of the scattering does not affect image quality. To further confirm sub-diffraction limit imaging in our PhC, we measured near-field intensity along the $\Gamma$-M axis of the crystal at $\lambda = 1489$ nm (Fig. 2c). At this wavelength an intensity peak is observed inside

the PhC slab (labelled "2"); such a mid-slab intensity peak further confirms the well-known characteristic of an effective negative index material [13,1]. The spatial FWHM (along $\underline{x}$-direction) of this second focus spot, inside the PhC slab, was analyzed to be ~0.96 ± 0.02 um.

These experimental features are also verified via rigorous 3D FDTD simulations (Figs. 2e and 2f). Negative refraction sub-diffraction imaging is optimized at a wavelength of $\lambda$ = 1461 nm. For comparison, the theoretical field profile at $\lambda$ = 1870 nm (in the first TM-like band, with positive index) has an image cross-section significantly larger than at 1461 nm. Our numerical computations suggest a sub-diffraction imaging resolution of 0.67 µm (0.46$\lambda$) is achieved at $\lambda$ = 1461 nm, matching within 2% from our NSOM measurements. This small deviation could be from fabrication-simulation mismatches or the finite impulse response of the NSOM fiber tip.

The wavelength (1489 nm) for the minimum image size is closely related to the resonant excitation of a surface slab mode near this wavelength for evanescent wave amplification. Fig. 3 presents the experimental evidence, where the intensity frequency dependence for three different locations at the PhC input/output facets is shown. Each location is averaged over ~3$a$ length along the crystal facet to eliminate local field variations. At *any* location along the slab facet, in addition to the three shown, the averaged transmitted signal has a maxima at 1489 nm. For wavelengths shorter than 1489 nm, the modes become leaky and the increased losses impair the imaging efficiency; for longer wavelengths, the frequency separation from the surface slab modes increases and thus recovery of the evanescent wave decreases. This behaviour has also been suggested in a recent theoretical study [10]. Anisotropy of the phase refractive index also increases with the wavelength, again reducing imaging efficiency.

Moreover, due to refocusing of the non-propagating waves, we note that local field intensity along the focus axis can be enhanced when on-resonance ($\lambda = 1489$ nm), compared to off-resonance, and albeit only at locations close to the focus axis. This is shown in Fig. 3c. At all other wavelengths (such as at the 1480 nm shown), there is less refocusing. Moreover, in comparison, a homogenous slab does not show the strong frequency dependence or refocusing of the evanescent waves. Moreover, our FDTD simulations likewise verify this local refocusing of the evanescent wave by the excited surface slab modes. We note that the local field (not energy flux, which is conserved) is the quantity measured by the NSOM.

The image formation quality is also affected by optical losses in the PhC slab. An averaged transmission loss of $0.59 \pm 0.003$ dB is found at $\lambda = 1500$ nm (near-resonance but not at 1489 nm, so as not to include the surface state enhancements) by analyzing optical intensity at any two transverse PhC slab locations. This loss converts to an estimated $\text{Im}(n)$ to be $\sim 4.4 \times 10^{-3}$ and correspondingly a FOM of $\sim 380$. Coupling losses from the waveguide into the PhC is significantly larger at $\sim 2.2$ dB, mainly due to impedance and mode mismatch. Nonetheless, the negative refraction FOM is about $125\times$ improvement over recent experimental efforts and more than $15\times$ better than best theoretical efforts [12], and the negative refraction PhC is scalable to optical frequencies using wide band-gap materials and smaller lattice constants.

To further verify the existence of an effective negative refractive index, we also fabricated a wedge-shaped PhC slab ($a = 443$ nm, $r = 135$ nm and $t = 320$ nm) and a homogeneous Si slab with identical geometry and dimensions. A negative refraction PhC slab would form an image on the opposite side of the normal [17], as compared to a positive refraction medium. Although not excited by a perfect plane-wave, the predominant direction of wavevectors excitation along the longitudinal axis of the waveguide provides an indication of negative refraction [18]. Fig. 4 shows two example near-field measurements, where the PhC slab forms an image at the "negative" side of the surface normal at the PhC-air interface (Fig. 4a; right inset) at $\lambda$ = 1545 nm. In contrast, the homogeneous slab (Fig. 4a; left inset) shows the image on the *opposite* side of the normal. The effective negative index is estimated ~ -1.29 at $\lambda$ = 1545 nm. A minimum image spatial FWHM of 2.2 ± 0.15 μm is observed at 1545 nm. Although this image is not beyond the diffraction limit, the minimum spatial FWHM in the wedge-shaped slab occurs at the *same* normalized frequency (a/$\lambda$) that corresponds with the earlier rectangular slab with different PhC *a* and *r* parameters (Fig. 2 and 3). Moreover, because the wedge-shape slab has different surface termination at the input and output facets, it does not support surface modes at the same frequency, and hence does not allow sub-diffraction limit imaging. In addition, we note that our PhC structures are not designed to support super-collimation recently observed [19], as evidenced in the band structure of Fig. 1b.

The fabrication disorder in the PhC slabs (shown in Fig. 1a) was statistically parameterized [20] with resulting hole radius 119.36 ± 2.29 nm, lattice period 433.50 ± 1.53 nm (~ 0.004*a*), ellipticity 6.29 nm ± 0.51 nm, and edge roughness correlation length of 27 nm. These small variations are below ~ 0.05*a* disorder theoretical target [21], to minimize distortions of our negative refraction image.

This first experimental observation of sub-diffraction imaging through bound surface states in negative refraction photonic crystals at the near-infrared suggests interesting opportunities for scaling to visible and shorter wavelengths for imaging, detection and nanolithography applications, where the commonly perceived diffraction limit is no longer a barrier. Further fundamental investigations include extension to higher-fold symmetry quasicrystals for non-near-field focusing regimes [22], zero-$\bar{n}$ superlattice metamaterials [23], as well as transformation optics [24] for manipulating the flow of light [25].

This work is supported in part by the National Science Foundation ECCS-0622069 (R. C. and C. W. W) and ECCS-0523386 (N. C. P. and R. M. O), the New York State Foundation for Science, Technology and Innovation, and the Columbia University Initiatives in Science and Engineering. We are particularly grateful for the review comments for further motivating new NSOM studies and results.

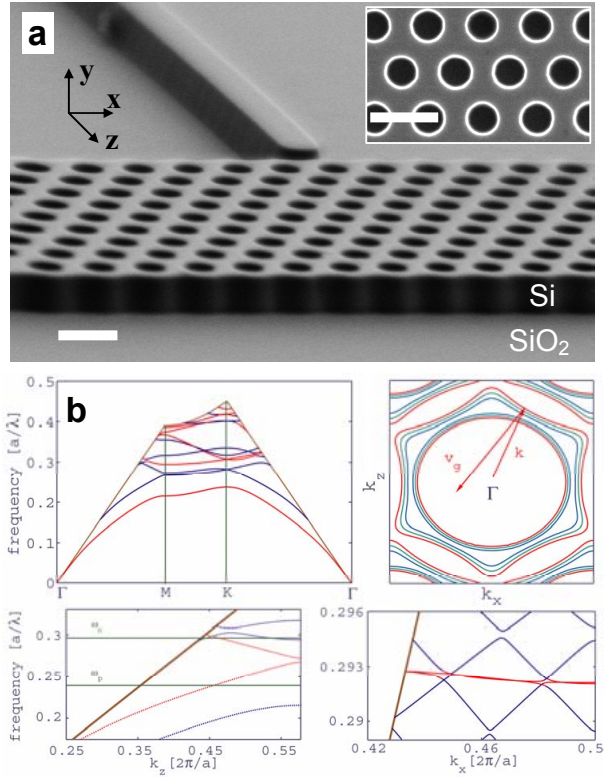

Fig. 1 (color online). (a) Scanning electronic micrograph of negative refraction photonic crystal metamaterial. Scale bar: 500 nm. (b) Photonic band structure of an air-hole hexagonal PhC slab with $r = 0.279a$; $t = 0.744a$ with $a = 430$ nm (top-left panel). The TE-like (TM-like) photonic bands are depicted in red (blue) and correspond to effective positive (negative) values of parity $\sigma_y$. Bottom-left panel: zoom-in of frequency range of experiments. Horizontal lines correspond to two frequencies at which the effective index of refraction is negative ($\omega_n$) or positive ($\omega_p$). Top-right panel: equi-frequency curves of the photonic band with negative refraction, computed for $\omega = 0.282$ (red), $\omega = 0.285$ (green), and $\omega = 0.288$ (blue). Bottom-right panel: zoom-in of band structure of PhC transverse guiding modes.

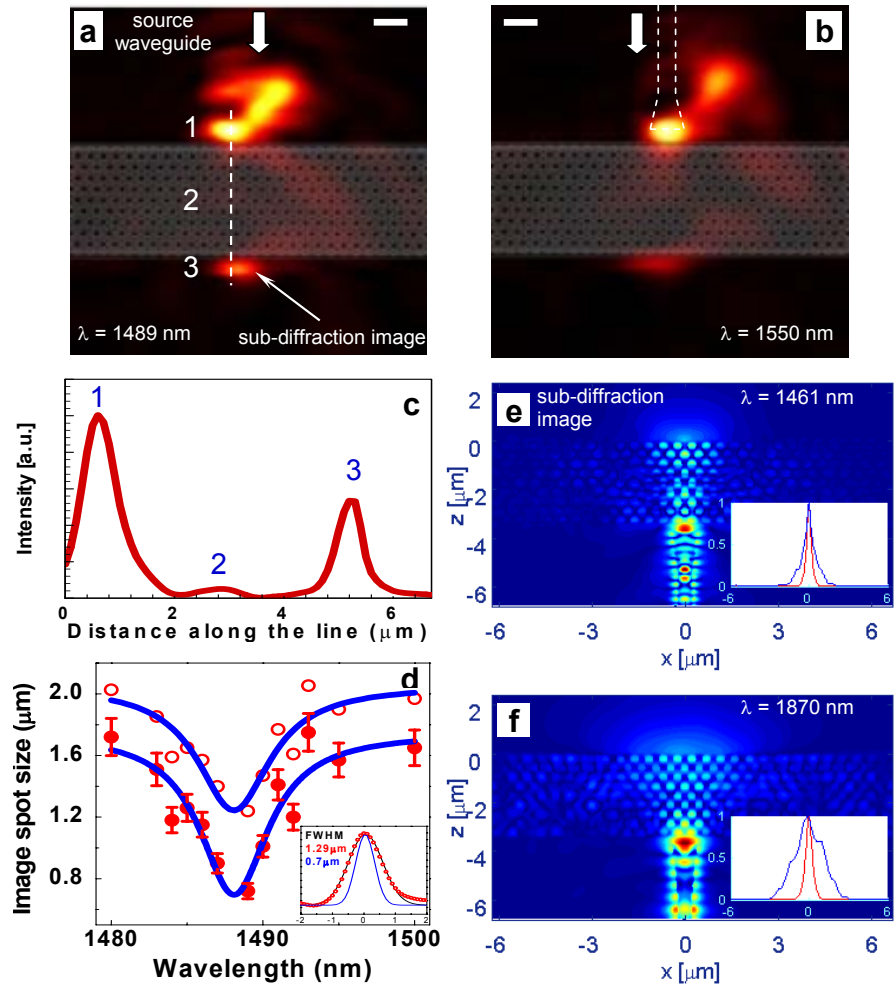

Fig. 2 (color online). (a) and (b) Near-field observation of near-infrared negative refraction sub-diffraction imaging. SEM image superimposed. Scale bar: 1 μm. (c) Near-field intensity profile along the Γ-M symmetry axis, at λ = 1489 nm. (d) Spectral dependence of FWHM image waist, before (empty circles) and after (filled dots) deconvolution. Inset: image intensity cross-section at 1489 nm, before (red dots) and after (blue curve) deconvolution, parallel to PhC surface. (e) and (f), FDTD results (*H*-field) of sub-diffraction imaging. Inset: intensity cross-section across the input spots (red curves) and the focused images (blue curves).

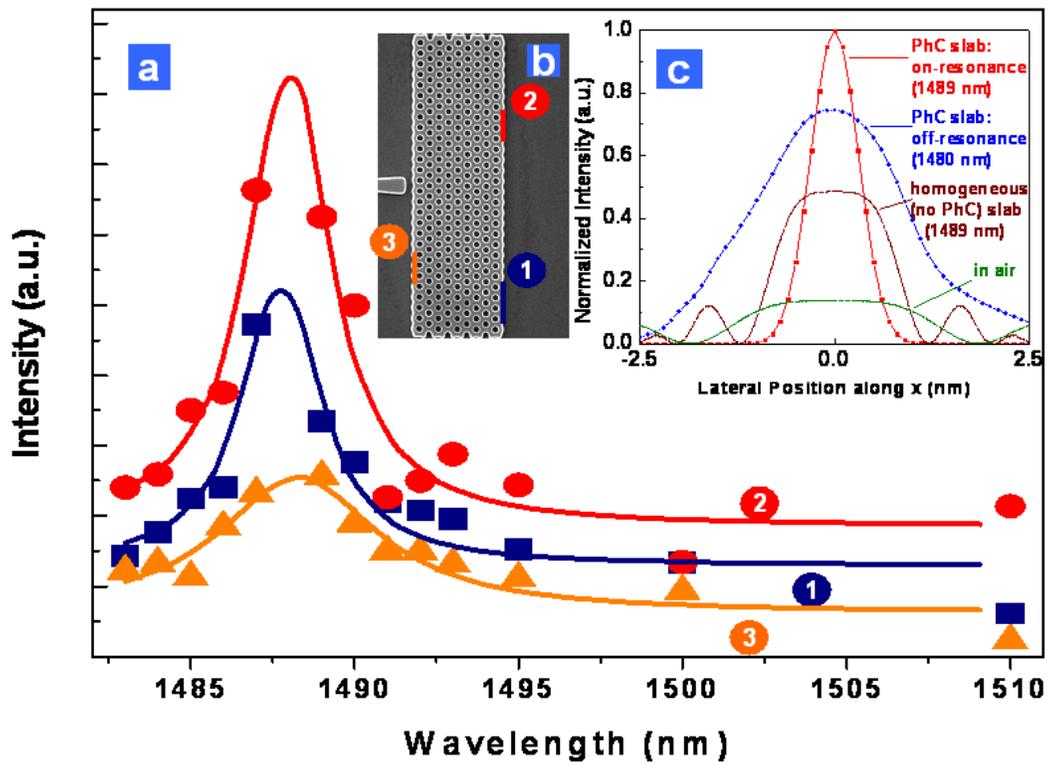

Fig. 3 (color online). Observation of resonant bound surface states excitation at wavelength of highest sub-diffraction limit imaging resolution. (a) Resonant wavelength dependence of input and output field intensities at three spatial domains. (b) locations of examined field intensities. (c) Refocusing of the near-field intensity, comparing local intensity at the output facet of the PhC slab (on resonance:1489 nm; off-resonance: 1480 nm), homogeneous (no PhC) slab (1489 nm), and in air.

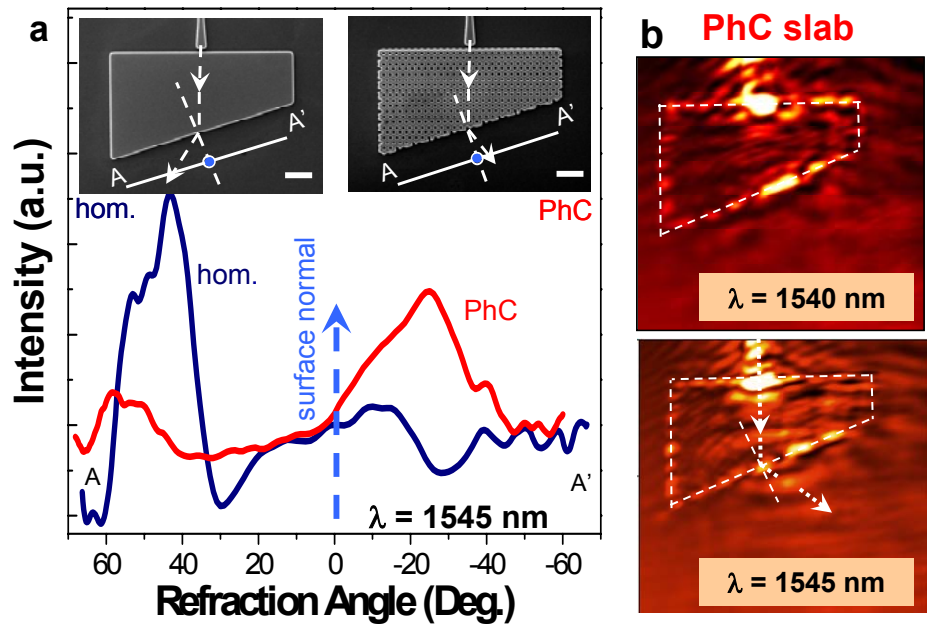

Fig. 4 (color online). Observation of negative refraction in wedge-like photonic crystal structure. (a) Field intensity measured along A-A'. Inset scale bar: 2 μm. (b) Near-field comparison of PhC slab showing spectral dependence.